\documentclass[twocolumn]{svjour3}          
\smartqed  
\usepackage{tabularx,xtab,array,float}
\usepackage{epsfig,epstopdf,wrapfig,psfrag}
\usepackage{amsmath,amssymb,graphicx,graphics}
\usepackage{tikz}
\usepackage{balance}
\usepackage{cite}
\usepackage{hyperref}
\makeatletter
\newcommand*{\encircled}[1]{\relax\ifmmode\mathpalette\@encircled@math{#1}\else\@encircled{#1}\fi}
\newcommand*{\@encircled@math}[2]{\@encircled{$\m@th#1#2$}}
\newcommand*{\@encircled}[1]{%
  \tikz[baseline,anchor=base]{\node[draw,circle,outer sep=0pt,inner sep=.2ex] {#1};}}
\makeatother

\renewcommand{\vec}[1]{\boldsymbol{\mathrm{#1}}}
\begin{document}

\title{Nonprehensile Manipulation of a Stick\\ Using Impulsive Forces}


\author{Aakash Khandelwal \and
        Nilay Kant \and
        Ranjan Mukherjee
}


\institute{A. Khandelwal \at
Department of Mechanical Engineering, Michigan State University, East Lansing, MI 48824, USA \\
\email{khande10@egr.msu.edu} 
\and
N. Kant \at
Department of Mechanical Engineering, Michigan State University, East Lansing, MI 48824, USA \\
\email{kantnila@msu.edu}
\and
R. Mukherjee \at
Department of Mechanical Engineering, Michigan State University, East Lansing, MI 48824, USA \\
\email{mukherji@egr.msu.edu}
}

\date{Received: date / Accepted: date}

\maketitle

\begin{abstract}
The problem of nonprehensile manipulation of a stick in three-dimensional space using intermittent impulsive forces is considered. The objective is to juggle the stick between a sequence of configurations that are rotationally symmetric about the vertical axis. The dynamics of the stick is described by five generalized coordinates and three control inputs. Between two consecutive configurations where impulsive inputs are applied, the dynamics is conveniently represented by a Poincar\'e map in the reference frame of the juggler. Stabilization of the orbit associated with a desired juggling motion is accomplished by stabilizing a fixed point on the Poincar\'e map. The Impulse Controlled Poincar\'e Map approach is used to stabilize the orbit, and numerical simulations are used to demonstrate convergence to the desired juggling motion from an arbitrary initial configuration. In the limiting case, where consecutive rotationally symmetric configurations are chosen arbitrarily close, it is shown that the dynamics reduces to that of steady precession of the stick on a hoop.\
\end{abstract}


\section*{Nomenclature}
\begin{tabularx}{\columnwidth}{lX}
$g$ &acceleration due to gravity, (m/s$^2$)\\$h_x$, $h_y$, $h_z$ &Cartesian coordinates of the center-of-mass of the stick in the $xyz$ frame, (m)\\
$\ell$ &length of the stick, (m)\\
$m$ &mass of the stick, (kg)\\
$r$ &distance of point of application of impulsive force from the center-of-mass of the stick, measured positive along the $z_2$ axis, (m)\\
$v_x$, $v_y$, $v_z$ &velocities of the center-of-mass of the stick in the $xyz$ frame, (m/s)\\
$xyz$ &inertial reference frame\\
$x_0y_0z_0$ &reference frame with origin at the center-of-mass of the stick; aligned with the $xyz$ frame\\
$x_1y_1z_1$ &reference frame with origin at the center-of-mass of the stick; obtained by rotating $x_0y_0z_0$ frame by $\alpha$ about the $z_0$ axis\\
$x_2y_2z_2$ &reference frame with origin at the center-of-mass of the stick; obtained by rotating $x_1y_1z_1$ frame by $\beta$ about the $y_1$ axis\\
$x_3y_3z_3$ &body-fixed reference frame with origin at the center-of-mass of the stick; obtained by rotating $x_2y_2z_2$ frame by $\gamma$ about the $z_2$ axis\\
$H_x$, $H_y$, $H_z$ &Components of the angular momentum about the center-of-mass of the stick in the $xyz$ frame, (kgm$^2$/s)\\
$I$ &non-negative magnitude of impulsive force applied on the stick, (Ns)
\end{tabularx}

\begin{tabularx}{\columnwidth}{lX}
$J$ &mass moment of inertia of the stick about the principal axes $x_3$ and $y_3$, (kgm$^2$)\\
$\vec R$ &rotation matrix that transforms a vector from the $x_3y_3z_3$ frame to the $x_0y_0z_0$ frame\\
$\alpha$, $\beta$, $\gamma$ &$zyz$ Euler angle sequence describing the stick orientation, (rad)\\
$\phi$ &angle that the impulsive force makes with the $x_2$ axis, measured positive about the $z_2$ axis, (rad)\\
$\dot\alpha$, $\dot\beta$, $\dot\gamma$ &$zyz$ Euler angle rates, (rad/s)\\
$\bar{[.]}$ &$[.]$ expressed in the reference frame of the juggler
\end{tabularx}

\section{Introduction} \label{sec1}
With robots expected to perform increasingly complex tasks, it is imperative that nonprehensile manipulation be included in their repertoire. Nonprehensile manipulation represents an important class of problems in which objects are manipulated without grasping; they are subjected to unilateral constraints and need not strictly follow the motion of the manipulator \cite{lynch1996dynamic, lynch1999dynamic, mason1999progress, lynch2003control, woodruff2017planning, ruggiero2018survey}. The advantages of nonprehensile manipulation over prehensile manipulation \cite{lynch1999dynamic, ruggiero2018survey} include:\

\begin{itemize}
\item added flexibility in manipulation since additional surfaces of the manipulator may be used to make unilateral contact with the object.
\item an increased workspace, exceeding the kinematic reach of the manipulator.
\item the ability to control more degrees-of-freedom (DOFs) than that of the manipulator.
\end{itemize}

Nonprehensile manipulation can be subdivided into tasks where the contact between the manipulator and the object is continuous, and tasks where the contact is intermittent. Continuous contact occurs, for example, when an object is pushed by a manipulator to slide or roll on a surface. Examples of the latter class of problems include the nonprehensile manipulation primitives of dynamic catching, throwing, and \emph{batting} \cite{ruggiero2018survey}, which combines dynamic catching and throwing into a single collision. Juggling is a nonprehensile manipulation task comprised of iterative batting primitives \cite{ruggiero2018survey}. The dynamics of juggling is hybrid (non-smooth), and involves the application of intermittent impulsive forces. The controllability and stability of hybrid dynamical systems has been investigated by many researchers \cite{brogliato2018feedback,brogliato2006on,goebel2012hybrid,lynch2001recurrence,tornambe1999modeling,zavalario2001direct}, but is not the focus of this paper.\

Juggling a point mass, such as a ball or a hockey puck, requires the application of a single impulsive force for each batting primitive to achieve a desired motion. The ball-juggling problem was investigated using a one-DOF table \cite{zavalario1999on} first, and later using a two-DOF manipulator \cite{brogliato2000on}. In these works, the impact rule was incorporated in the dynamic model, comprised of both the ball and the robot. \emph{Blind-jugglers}, which juggle balls without relying on external sensors, have been analysed in \cite{ronsse2007rhythmic, reist2012design}, and other solutions to juggling of point masses have been proposed in \cite{sanfelice2007hybrid, schaal1993open, spong2001impact}. Compared to a point mass, which is described by position coordinates only, a stick represents an extended object that is described by both position and orientation coordinates. Therefore, both the physical task and the mathematical problem of stick-juggling are more challenging than juggling a ball or a point mass.\

Previous work on nonprehensile manipulation of sticks has documented the use of continuous-time inputs for rotary \emph{propeller} motion \cite{nakaura2004enduring, shiriaev2006generating, nakamura2009enduring} but there is limited literature on manipulation of sticks using impulsive forces. For nonprehensile manipulation of a stick using impulsive forces, the location, direction, and magnitude of the impulsive force has to be taken into consideration for every batting primitive. An open-loop strategy for planar stick-juggling was presented by Schaal and Atkeson \cite{schaal1993open}; the dynamics of the system was however not completely modeled. The complete dynamic model and closed-loop control designs for planar stick-juggling were presented by Kant and Mukherjee in \cite{kant2021non, kant2022juggling}. Nonprehensile manipulation of a stick in three-dimensional space has not appeared in the literature and this work is the first to present a hybrid dynamic model that lends itself to closed-loop control design. The dynamic model represents an underactuated system with five generalized coordinates and three control inputs; the three-dimensional juggling problem is more challenging than the planar case where the stick is described by three generalized coordinates and two control inputs. The Impulse Controlled Poincar\'e Map (ICPM) approach \cite{kant2020orbital,kant2022juggling}, in which impulsive forces are intermittently applied on a Poincar\'e section, is used for stabilizing the hybrid orbit that describes the desired juggling motion.\

This paper is organized as follows. The problem of juggling a stick in three-dimensional space using impulsive forces is formally stated in section \ref{sec2}. The impulsive and continuous-time dynamics of the stick is presented in section \ref{sec3}. The hybrid dynamics of the stick between two consecutive configurations where impulsive inputs are applied is described with the help of a nonlinear discrete-time Poincar\'e map in section \ref{sec4}. The control design is presented in section \ref{sec5}. Simulation results are presented in section \ref{sec6}. Section \ref{sec7} considers the special cases of planar juggling and steady precession. It is shown that previously derived expressions for planar symmetric juggling \cite{kant2021non, kant2022juggling} can be recovered from the current, more general, formulation. The steady precession of the stick is shown to be a limiting case of the hybrid dynamics. Concluding remarks are provided in section \ref{sec8}.\

\section{Problem Description} \label{sec2}
\begin{figure}[b!]
	\centering
	\psfrag{A}[][]{\small{$x$}}
	\psfrag{B}[][]{\small{$y$}}
	\psfrag{C}[][]{\small{$z$}}
	\psfrag{D}[][]{\small{$x_0$}}
	\psfrag{E}[][]{\small{$y_0$}}
	\psfrag{F}[][]{\small{$z_0$, $z_1$}}
	\psfrag{G}[][]{\small{$x_1$}}
	\psfrag{H}[][]{\small{$y_1$, $y_2$}}
	\psfrag{I}[][]{\small{$x_2$}}
	\psfrag{J}[][]{{$z_2$, $z_3$}}
	\psfrag{L}[][]{\small{$\alpha$}}
	\psfrag{M}[][]{\small{$\beta$}}
	\psfrag{N}[][]{\small{$g$}}
	\psfrag{P}[][]{\small{$\gamma$}}
	\psfrag{S}[][]{\small{$x_3$}}
	\psfrag{T}[][]{\small{$y_3$}}
	\psfrag{Z}[][]{\small{$G$}}
	\includegraphics[width=0.73\hsize]{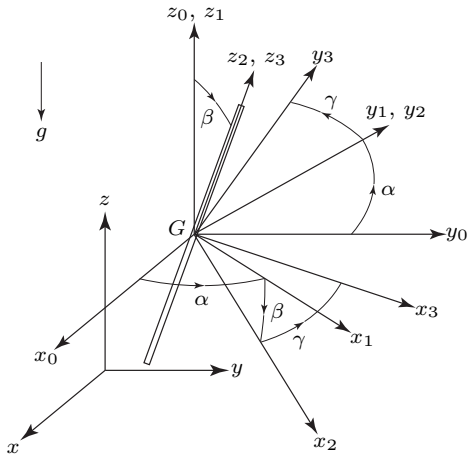}
	\caption{A rigid stick in the three-dimensional space has six DOF and is described by the configuration variables $(h_x, h_y, h_z, \alpha, \beta, \gamma)$.}
	\label{Fig1}
\end{figure}
\begin{figure}[b!]
	\centering
	\psfrag{A}[][]{\small{$\beta^*$}}
	\psfrag{B}[][]{\small{$\Delta\alpha^*$}}
	\psfrag{C}[][]{\small{$z_0, z_1$}}
	\psfrag{D}[][]{\small{$z_2$}}
	\psfrag{E}[][]{\small{$x_1$}}
	\psfrag{F}[][]{\small{$y_1, y_2$}}
	\psfrag{G}[][]{\small{$z$}}
	\psfrag{H}[][]{\small{$x_2$}}
	\psfrag{Z}[][]{\small{$G$}}
	\psfrag{N}[][]{\small{$g$}}
	\includegraphics[width=0.51\hsize]{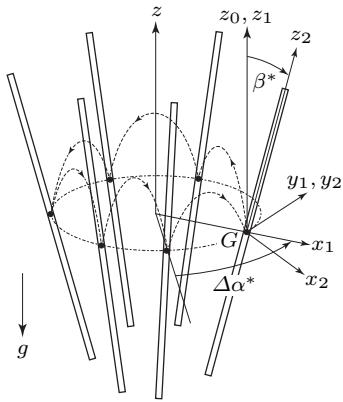}
	\caption{A sequence of configurations of the stick at steady-state, rotationally symmetric about the $z$ axis.}
	\label{Fig2}
\end{figure}
\begin{figure}[t!]
	\centering
	\psfrag{A}[][]{\small{$\phi$}}
	\psfrag{B}[][]{\small{$r$}}
	\psfrag{C}[][]{\small{$I$}}
	\psfrag{D}[][]{\small{$z_2$}}
	\psfrag{F}[][]{\small{$y_2$}}
	\psfrag{H}[][]{\small{$x_2$}}
	\psfrag{Z}[][]{\small{$G$}}
	\psfrag{G}[][]{\small{$z_0, z_1$}}
	\psfrag{E}[][]{\small{$\beta^*$}}
	\psfrag{N}[][]{\small{$g$}}
	\includegraphics[width=0.28\hsize]{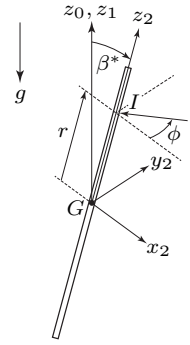}
	\caption{Control inputs $(I, r, \phi)$: the impulsive force is applied when $\beta = \beta^*$.}
	\label{Fig3}
\end{figure}

Consider the six DOF stick shown in Fig.\ref{Fig1}, which can move freely in three-dimensional space. The center-of-mass of the stick is denoted by $G$. The configuration of the stick is specified by the generalized coordinates $(h_x, h_y, h_z, \alpha, \beta, \gamma)$. The stick is assumed to be symmetric about the body-fixed $z_3$ axis, which is the same as the $z_2$ axis. Due to this axisymmetry, the rotation by $\gamma$ about the $z_2$ axis is imperceptible. The objective is to juggle the stick between a sequence of configurations which, at steady-state, are rotationally symmetric about the inertial $z$ axis - see Fig.\ref{Fig2}. Each configuration in the sequence satisfies $\beta = \beta^*$, $\beta^* \in (0, \pi/2)$,  and can be obtained from the previous configuration by a fixed change in $\alpha$, equal to $\Delta\alpha^*$. For the sequence of configurations, the coordinates of $G$ lie on a circle parallel to the $xy$ plane. The stick is juggled using purely impulsive forces applied normal to the stick; the impulsive forces are applied only when $\beta = \beta^*$. Since the impulsive force is applied normal to the stick, the impulsive force lies on a plane parallel to the $x_2y_2$ plane - see Fig.\ref{Fig3}. The control inputs are the triplet $(I, r, \phi)$; at steady-state, they assume the constant values $(I^*, r^*, \phi^*)$.\

\section{System Dynamics} \label{sec3}

\subsection{Coordinate Transformations} \label{sec31}

The elementary rotation matrices describing a rotation about the $y$ and $z$ axes by angle $\theta$ are given as:
\begin{equation} \label{eq1}
	\vec R_y(\theta)\! =\! \setlength\arraycolsep{2.5pt}\begin{bmatrix}
		\cos\theta &0 &\sin\theta \\
		0 &1 &0 \\
		-\sin\theta &0 &\cos\theta \\
	\end{bmatrix}, \,\,
	\vec R_z(\theta)\! =\! \setlength\arraycolsep{2.5pt}\begin{bmatrix}
		\cos\theta &-\sin\theta &0 \\
		\sin\theta & \cos\theta &0 \\
		0 &0 &1 \\
	\end{bmatrix}
\end{equation}

\noindent A vector $\vec{p}$ described in the inertial reference frame can be related to the same vector expressed in the body-fixed frame by the expression
\begin{equation} \label{eq2}
\vec{p} = \vec R \vec{p}_{\rm b}, \qquad \vec R \triangleq \vec R_z(\alpha) \vec R_y(\beta) \vec R_z(\gamma)
\end{equation}

\noindent We define the position and velocity vectors in the inertial frame
\begin{equation} \label{eq3}
\vec{h} \triangleq \begin{bmatrix} h_x \\ h_y \\ h_z \end{bmatrix}, \quad \vec{v} \triangleq \begin{bmatrix} v_x \\ v_y \\ v_z \end{bmatrix} = \begin{bmatrix} \dot h_x \\ \dot h_y \\ \dot h_z \end{bmatrix}
\end{equation}

\noindent The vector of Euler angle rates can be related to the angular velocity vector in the inertial frame using the relationship
\begin{equation} \label{eq4}
\vec{\omega} = \vec S \vec \Omega, \,\,
\vec S \triangleq \setlength\arraycolsep{2.5pt}\begin{bmatrix}
		0 &-\sin\alpha &\cos\alpha\sin\beta \\
		0 &\cos\alpha &\sin\alpha\sin\beta \\
		1 &0 &\cos\beta
	\end{bmatrix}, \,\,
\vec \Omega \triangleq \begin{bmatrix} \dot\alpha \\ \dot\beta \\ \dot\gamma \end{bmatrix}
\end{equation}

\noindent Assuming a slender stick, the moment of inertia matrix $\vec J$ in the inertial reference frame is related to the inertia matrix $\vec J_{\rm b}$ in the body-fixed $x_3y_3z_3$ frame by the expression
\begin{equation} \label{eq5}
	\vec J = \vec R \vec J_{\rm b} \vec R^T, \qquad \vec J_{\rm b} \triangleq \mathrm{diag}\begin{bmatrix} J & J & 0\end{bmatrix}
\end{equation}

\noindent where the mass moment of inertia about the principal axis $z_3$ is zero. The matrix $\vec J$ is found to be independent of $\gamma$; this is due to the symmetry of the stick about the $z_3$ axis. Using \eqref{eq4}, the angular momentum of the stick in the inertial reference frame, $\vec H \triangleq \begin{bmatrix} H_x & H_y & H_z \end{bmatrix}^T$, can be expressed as
\begin{equation} \label{eq6}
\vec{H} = \vec J \vec{\omega} = \vec J \vec S \vec\Omega
\end{equation}

\noindent where
\begin{equation} \label{eq7}
\vec J \vec S = J \begin{bmatrix}
-\cos\alpha \sin\beta \cos\beta & -\sin\alpha & 0 \\
-\sin\alpha \sin\beta \cos\beta & \cos\alpha & 0 \\
\sin^2\beta & 0 & 0
\end{bmatrix}
\end{equation}

\noindent The above expression, which was obtained using \eqref{eq4} and \eqref{eq5}, indicates that the matrix $\vec J \vec S$ is singular and $\vec H$ is a function of $\dot\alpha$ and $\dot\beta$, but not $\dot\gamma$.\

\subsection{Impulsive Dynamics} \label{sec32}
Let $t_k,\quad k = 1, 2, \cdots,$ denote the instants of time when impulsive inputs are applied on the stick. Furthermore, let $t_k^-$ and $t_k^+$ denote the instants of time immediately before and after application of the impulsive inputs. Since impulsive inputs cause no change in position coordinates \cite{brogliato1999nonsmooth, kant2019estimation}, we have
\begin{equation*}
\vec{h}(t_k^+) = \vec{h}(t_k^-), \,\, \alpha(t_k^+) = \alpha(t_k^-) \triangleq \alpha_k, \,\, \beta(t_k^+) = \beta(t_k^-) = \beta^*
\end{equation*}

\noindent where $\alpha_k$ is defined for notational simplicity and $\beta(t_k^+) = \beta(t_k^-) = \beta^* \, \forall \, k$ follows from our discussion in section \ref{sec2}. At time $t_k$, the vector of impulsive force can be written in the inertial reference frame as
\begin{equation} \label{eq8}
\begin{split}	
\vec{I}_k &= \vec R_z(\alpha_k) \vec R_y(\beta^*)\begin{bmatrix} -I_k \cos\phi_k \\ -I_k \sin\phi_k \\ 0 \end{bmatrix} = I_k \vec f_k, \\
\vec f_k &\triangleq \begin{bmatrix}
\quad \sin\alpha_k \sin\phi_k - \cos\alpha_k \cos\beta^* \cos\phi_k\\
-\cos\alpha_k \sin\phi_k - \sin\alpha_k \cos\beta^* \cos\phi_k\\
\sin\beta^*\cos\phi_k
\end{bmatrix}
\end{split}
\end{equation} 

\noindent where $I_k = I(t_k)$ and $\phi_k = \phi(t_k)$. The vector from the center-of-mass $G$ to the point of application of the impulsive force can be written in the inertial reference frame as
\begin{equation} \label{eq9}
\vec{r}_k = \vec R_z(\alpha_k) \vec R_y(\beta^*)\begin{bmatrix} 0 \\ 0 \\ r_k \end{bmatrix}
= r_k \begin{bmatrix}
\cos\alpha_k \sin\beta^* \\
\sin\alpha_k \sin\beta^* \\
\cos\beta^*
\end{bmatrix}
\end{equation} 

\noindent where $r_k = r(t_k)$. The discontinuous jumps in velocities due to the impulsive inputs can be obtained from the linear and angular impulse-momentum relationships. The linear impulse-momentum relationship is given by
\begin{equation} \label{eq10}
m\vec{v}(t_k^+) = m\vec{v}(t_k^-) + \vec{I}_k
\end{equation}
which, using \eqref{eq8}, can be expressed as
\begin{equation} \label{eq11}
\vec{v}(t_k^+) = \vec{v}(t_k^-) + \frac{I_k}{m}\, \vec f_k
\end{equation}

\noindent The angular impulse-momentum relationship is given by
\begin{equation} \label{eq12}
\vec{H}(t_k^+) = \vec{H}(t_k^-) + \vec{r}_k \times \vec{I}_k
\end{equation}
Using \eqref{eq6}, \eqref{eq8}, and \eqref{eq9}, we get
\begin{equation} \label{eq13}
\begin{split}
\vec J_k \vec S_k \vec\Omega(t_k^+) = &\,\,\vec J_k \vec S_k \vec\Omega(t_k^-) \\ + &I_k r_k\! \begin{bmatrix}
\quad \sin\alpha_k \cos\phi_k + \cos\alpha_k \cos\beta^* \sin\phi_k \\
-\cos\alpha_k \cos\phi_k + \sin\alpha_k \cos\beta^* \sin\phi_k \\
-\sin\beta^*\sin\phi_k
\end{bmatrix}
\end{split}
\end{equation}

\noindent where $\vec J_k \vec S_k = \vec J(t_k) \vec S(t_k)$, and is obtained from \eqref{eq7} by using $\alpha = \alpha_k$ and $\beta = \beta^*$. It can be shown that two of the three equations in \eqref{eq13} are independent; these two equations can be used to solve for the two unknowns, namely,
\begin{subequations} \label{eq14}
\begin{align}
\dot \alpha(t_k^+) &= \dot \alpha(t_k^-) - \frac{I_k r_k \sin\phi_k}{J \sin\beta^*} \label{eq14a} \\
\dot \beta(t_k^+) &= \dot \beta(t_k^-) - \frac{I_k r_k \cos\phi_k}{J} \label{eq14b}
\end{align}
\end{subequations}

\begin{remark} \label{rem1}
The equations describing the impulsive dynamics are independent of $\gamma$ and $\dot\gamma$.\
\end{remark}

\subsection{Continuous-time Dynamics} \label{sec33}
Over the interval $t \in [t_k^+, t_{k+1}^-]$, the stick undergoes torque-free motion under gravity. The motion of the center-of-mass is therefore described by the differential equation\
\begin{equation} \label{eq15}
\dot{\vec h} = \vec v, \qquad \dot{\vec v} = \begin{bmatrix} 0 & 0 & -g \end{bmatrix}^T
\end{equation}
Using $\vec h(t_k^+) $ and $\vec v(t_k^+)$ as the initial conditions for $\vec h$ and $\vec v$, the solution to \eqref{eq15} is obtained as
\begin{subequations}\label{eq16}
\begin{align}
\vec h(t_{k+1}^-) &= \vec h(t_k^-) + \vec v(t_k^-)\delta_k + \frac{I_k\delta_k}{m} \vec f_k -\frac{1}{2} \begin{bmatrix} 0 \\ 0 \\ g\delta_k^2 \ \end{bmatrix} \label{eq16a} \\
\vec v(t_{k+1}^-) &= \vec v(t_k^-) + \frac{I_k}{m} \vec f_k - \begin{bmatrix} 0 \\ 0 \\ g\delta_k \ \end{bmatrix} \label{eq16b}
\end{align}
\end{subequations}
where $\delta_k \triangleq (t_{k+1}^- - t_k^-)$ is the interval between the $k$-th and $(k+1)$-th impulsive input; it is also the duration of the flight phase. Due to conservation of angular momentum, the rotational dynamics is described by the relation
\begin{align}
\vec H = \vec H(t_k^+) \quad \Rightarrow \quad \vec J \vec S \vec\Omega = \vec H(t_k^+) \label{eq17}
\end{align}

\noindent Since $\vec J \vec S$ is singular, \eqref{eq17} yields the following three equations that are not all independent
\begin{subequations} \label{eq18}
\begin{align}
J \dot\alpha \sin\beta \cos\beta &= - H_x(t_k^+) \cos\alpha - H_y(t_k^+) \sin\alpha \label{eq18a} \\
J \dot\beta &= - H_x(t_k^+) \sin\alpha + H_y(t_k^+) \cos\alpha \label{eq18b} \\
J \dot\alpha \sin^2\beta &= \,\,\,\, H_z(t_k^+) \label{eq18c}
\end{align}
\end{subequations}
where $H_x(t_k^+)$, $H_y(t_k^+)$, and $H_z(t_k^+)$ are the components of $\vec{H}(t_k^+)$, which can be obtained from \eqref{eq12}. We now solve for $\alpha$, $\dot\alpha$ and $\dot\beta$ at
$t_{k+1}^-$, knowing that $\beta = \beta^*$ at $t_{k+1}^-$, which is the end of the flight phase. Eliminating $\dot\alpha$ between \eqref{eq18a} and \eqref{eq18c} allows us to solve for $\alpha_{k+1}$ as
\begin{equation} \label{eq19}
\begin{split}
\alpha_{k+1} = &\alpha_k + \pi \\
&+ 2\arctan\!\!\left[\sin\beta^*\! \cos\beta^* \frac{\dot\alpha(t_k^-) - \dfrac{I_k r_k \sin\phi_k}{J \sin\beta^*}}{\dot\beta(t_k^-) - \dfrac{I_k r_k \cos\phi_k}{J}} \right]    
\end{split}
\end{equation}
The above relation holds under the constraint 
\begin{equation*}
\dot\beta(t_k^+) = \dot\beta(t_k^-) - \frac{I_k r_k \cos\phi_k}{J} < 0
\end{equation*}

\noindent Indeed, the physics of the problem requires $\dot\beta$ to be negative immediately following the application of each impulsive input. Equation \eqref{eq18c} can be solved for $\dot\alpha(t_{k+1}^-)$ to obtain
\begin{equation} \label{eq20}
\dot\alpha(t_{k+1}^-) = \dot \alpha(t_k^-) - \frac{I_k r_k \sin\phi_k}{J \sin\beta^*}
\end{equation}

\noindent Differentiating \eqref{eq18b} with respect to time and eliminating terms in $\alpha$ and $\dot\alpha$ using \eqref{eq18a} and \eqref{eq18c} allows us to obtain a decoupled second-order differential equation in $\beta$:
\begin{align} 
J \ddot\beta &= H_z^2(t_k^+) \cot\beta \csc^2\beta \notag \\
\Rightarrow\quad J \dot\beta d\dot\beta &= H_z^2(t_k^+) \cot\beta \csc^2\beta d\beta \label{eq21}
\end{align}

\noindent The above equation can integrated subject to initial conditions $\dot\beta = \dot\beta(t_k^+)$ and $\beta = \beta^*$ to obtain
\begin{equation} \label{eq22}
\dot\beta^2 = - K_1 \cot^2\beta + K_2
\end{equation}

\noindent where $K_1$ and $K_2$ are constants over the duration of the flight phase and are given by the relations\
\begin{equation*}
\begin{split}
K_1 &\triangleq \sin^4\!\beta^*\left[\dot \alpha(t_k^-) - \frac{I_k r_k \sin\phi_k}{J \sin\beta^*} \right]^2 \\
K_2 &\triangleq \sin^2\!\beta^* \cos^2\!\beta^*\left[\dot \alpha(t_k^-) - \frac{I_k r_k \sin\phi_k}{J \sin\beta^*} \right]^2 \\ &\qquad\quad + \left[\dot \beta(t_k^-) - \frac{I_k r_k \cos\phi_k}{J} \right]^2
\end{split}
\end{equation*}

\noindent To obtain $\dot\beta(t_{k+1}^-)$, we substitute $\beta = \beta^*$ in \eqref{eq22} and simplify. Since $\beta = \beta^*$ both at the beginning and end of the flight phase, we get two solutions. One solution gives the value of $\dot\beta$ at the beginning of the flight phase and is identical to \eqref{eq14b}; the other gives the value of $\dot\beta$ at the end of the flight phase:
\begin{equation} \label{eq23}
\dot\beta(t_{k+1}^-) = - \dot \beta(t_k^-) + \frac{I_k r_k \cos\phi_k}{J}
\end{equation}

\noindent To obtain the minimum value of $\beta$ during the flight phase, we substitute $\dot\beta = 0$ in \eqref{eq22}; this yields:
\begin{equation} \label{eq24}
\beta_{\rm min} = {\rm arccot}\sqrt{K_2/K_1}
\end{equation}

\noindent Separating the variables $\beta$ and $t$ in \eqref{eq22}, and observing that $\dot\beta$ is necessarily negative (positive) when $\beta$ changes from $\beta^*$  to $\beta_{\rm min}$
($\beta_{\rm min}$ to $\beta^*$), we obtain the time of flight $\delta_k$ as
\begin{equation} \label{eq25}
\begin{split}
\delta_k &= \int^{\beta_{\rm min}}_{\beta^*} \! \frac{d\beta}{-\sqrt{K_2 - K_1 \cot^2\beta}} + \int_{\beta_{\rm min}}^{\beta^*} \! \frac{d\beta}{\sqrt{K_2 - K_1 \cot^2\beta}} \\
&= 2\int_{\beta_{\rm min}}^{\beta^*} \! \frac{d\beta}{\sqrt{K_2 - K_1 \cot^2\beta}} \\
&= \frac{1}{\sqrt{K_1 + K_2}} \left[ \pi - 2\arctan\left( \frac{\sqrt{K_1 + K_2} \cot\beta^*}{\sqrt{K_2 - K_1 \cot^2\!\beta^*}} \right) \right]
\end{split}
\end{equation}

\begin{remark} \label{rem2}
Similar to the impulsive dynamics, the equations describing the continuous-time dynamics are independent of $\gamma$ and $\dot\gamma$. Therefore, the configuration of the stick can be adequately described by the five generalized coordinates $(h_x, h_y, h_z, \alpha, \beta)$.
\end{remark}

\section{Hybrid Dynamic Model}\label{sec4}
\subsection{Poincar\'e Map in Inertial Reference Frame}\label{sec41}

In light of remark \ref{rem2}, the generalized coordinates and their derivatives can be chosen to be the components of the state vector:
\begin{equation*}
X = \begin{bmatrix} h_x &h_y &h_z &v_x &v_y &v_z &\alpha &\beta &\dot\alpha &\dot\beta \end{bmatrix}^T
\end{equation*}

\noindent The vector of control inputs is denoted by
\begin{equation*}
U = \begin{bmatrix} I &r &\phi \end{bmatrix}^T
\end{equation*}

\noindent We now define the Poincar\'e section\footnote{It is assumed that the initial conditions of the stick are such that its trajectory intersects the Poincar\'e section before the first impulsive input is applied.} based on the configuration where the impulsive inputs are applied:
\begin{equation} \label{eq26}
S : \{X \in \mathbb{R}^{10} \,\, | \,\, \beta = \beta^* \}
\end{equation}

\noindent A point on $S$ can be described by the vector $Y$, $Y\! \subset \!X$, where
\begin{equation*}
Y = \begin{bmatrix} h_x &h_y &h_z &v_x &v_y &v_z &\alpha &\dot\alpha &\dot\beta \end{bmatrix}^T 
\end{equation*}

\noindent Using \eqref{eq16}, \eqref{eq19}, \eqref{eq20} and \eqref{eq23}, the hybrid dynamics, comprised of the impulsive and continuous-time dynamics, can be described by the map $\mathbb{P} : S \rightarrow S$ as follows:\

\begin{equation} \label{eq27}
\begin{split}
Y(t_{k+1}^-) &\!= \mathbb{P}\!\left[Y(t_k^-), U_k\right] \\
&\!=\! 
\begingroup
\renewcommand*{\arraystretch}{2.6}
\begin{bmatrix}
\begin{split}
h_x(t_k^-) \!+\! v_x(t_k^-)\delta_k &\!+\! \dfrac{I_k\delta_k}{m}(\sin\alpha_k \sin\phi_k \\&\!-\! \cos\alpha_k \cos\beta^* \cos\phi_k)
\end{split} \\
\begin{split}
h_y(t_k^-) \!+\! v_y(t_k^-)\delta_k &\!+\! \dfrac{I_k\delta_k}{m}(-\cos\alpha_k \sin\phi_k \\&\!-\! \sin\alpha_k \cos\beta^* \cos\phi_k)
\end{split} \\
h_z(t_k^-) \!+\! v_z(t_k^-)\delta_k \!+\! \dfrac{I_k\delta_k}{m}\sin\beta^*\cos\phi_k \!-\! \dfrac{1}{2}g\delta_k^2 \\
\begin{split}
v_x(t_k^-) &\!+\! \dfrac{I_k}{m}(\sin\alpha_k \sin\phi_k \\&\!-\! \cos\alpha_k \cos\beta^* \cos\phi_k)
\end{split} \\
\begin{split}
v_y(t_k^-) &\!+\! \dfrac{I_k}{m}(-\cos\alpha_k \sin\phi_k \\&\!-\! \sin\alpha_k \cos\beta^* \cos\phi_k)
\end{split} \\
v_z(t_k^-) + \dfrac{I_k}{m}\sin\beta^*\cos\phi_k  \!-\! g\delta_k \\
\begin{split} 
\alpha_k + \pi \qquad\qquad\qquad\qquad\\
+ 2\arctan\!\!\left[\sin\beta^*\! \cos\beta^* 
\dfrac{\dot\alpha(t_k^-) \!-\! \dfrac{I_k r_k \sin\phi_k}{J \sin\beta^*}}{\dot\beta(t_k^-) \!-\! \dfrac{I_k r_k \cos\phi_k}{J}} \right]
\end{split} \\
\dot \alpha(t_k^-) \!-\! \dfrac{I_k r_k \sin\phi_k}{J \sin\beta^*} \\
- \dot \beta(t_k^-) \!+\! \dfrac{I_k r_k \cos\phi_k}{J}
\end{bmatrix}
\endgroup
\end{split}
\end{equation}

\begin{figure}[h]
	\centering
	\psfrag{A}[][]{\small{$S$}}
	\psfrag{B}[][]{\scriptsize{$\encircled{2}$}}
	\psfrag{F}[][]{\scriptsize{$\encircled{1}$}}
	\psfrag{C}[][]{\scriptsize{$\encircled{3}$}}
	\psfrag{E}[][]{\scriptsize{$\encircled{4}$}}
	\psfrag{H}[][]{\scriptsize{$\encircled{5}$}}
	\psfrag{G}[][]{\scriptsize{$\encircled{6}$}}
	\psfrag{K}[][]{\scriptsize{
\begin{tabular}{l}
$\encircled{1}: Y(t_{k}^-)$\\ [0.5ex]
$\encircled{2}: Y(t_{k}^+)$\\ [0.5ex]
$\encircled{3}: Y(t_{k+1}^-)$\\ [0.5ex]
$\encircled{4}: Y(t_{k+1}^+)$\\ [0.5ex]
$\encircled{5}: Y(t_{k+2}^-)$ \\ [0.5ex]
$\encircled{6}: Y(t_{k+2}^+)$ \\
\end{tabular}}}
\includegraphics[width=0.64\hsize]{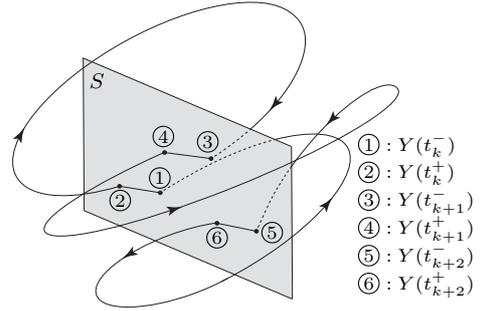}
\caption{Steady-state trajectory of the stick in the inertial reference frame for the particular case of $\Delta\alpha^* = 2\pi/3$; starting from the configuration $\protect\encircled{1}: Y(t_{k}^-)$, the stick returns to this configuration at $t_{k+3}^-$ following the application of three control actions at $t_k$, $t_{k+1}$, and $t_{k+2}$.}
\label{Fig8}
\end{figure}

\begin{remark} \label{rem3}
All elements of the vector $Y$ in \eqref{eq27} are defined in the inertial reference frame.
\end{remark}

\begin{remark} \label{rem4}
The objective is to juggle the stick between a sequence of configurations which are rotationally symmetric about the $z$ axis at steady-state - see problem description in section \ref{sec2}. These steady-state configurations do not correspond to a fixed point of the map $\mathbb{P}$ in \eqref{eq27} since each variable in the set $\{h_x, h_y, v_x, v_y, \alpha\}$ does not converge to a fixed value. This is illustrated with the help of Fig.\ref{Fig8}, which shows the Poincar\'e section $S$ and the steady-state trajectory of the stick for the particular case of $\Delta\alpha^* = 2\pi/3$. The stick is juggled between three configurations that are rotationally symmetric about the vertical axis and three control actions are required to bring the stick back to the original configuration, \emph{i.e.}, $Y(t_{k+3}^-)=Y(t_{k}^-)$. The impulsive dynamics of the stick is shown by the line segments $\encircled{1}\!\rightarrow\!\encircled{2}$, $\encircled{3}\!\rightarrow\!\encircled{4}$, and $\encircled{5}\!\rightarrow\!\encircled{6}$ on $S$; the continuous-time dynamics is shown by the trajectories
$\encircled{2}\!\rightarrow\!\encircled{3}$, $\encircled{4}\!\rightarrow\!\encircled{5}$, and $\encircled{6}\!\rightarrow\!\encircled{1}$. It can be seen that the steady-state configurations $Y(t_{k}^-)$, $Y(t_{k+1}^-)$, and $Y(t_{k+2}^-)$ are different in the inertial coordinate system and do not correspond to a fixed point of $\mathbb{P}$. This problem can be alleviated by defining the map in a rotating reference frame following the approach in \cite{kant2021non}.\
\end{remark}

\subsection{Poincar\'e Map in the Reference Frame of the Juggler}\label{sec42}

\begin{figure}[h]
	\centering
	\psfrag{A}[][]{\small{$x$}}
	\psfrag{B}[][]{\small{$y$}}
	\psfrag{C}[][]{\small{$z$}}
	\psfrag{D}[][]{\small{$z_2$}}
	\psfrag{F}[][]{\small{$O$}}
	\psfrag{H}[][]{\small{$P(t_k)$}}
	\psfrag{L}[][]{\small{$P(t_{k+1})$}}
	\psfrag{Z}[][]{\small{$G$}}
	\psfrag{G}[][]{\small{$z_1$}}
	\psfrag{E}[][]{\small{$\beta^*$}}
	\psfrag{M}[][]{\small{$x_1$}}
	\psfrag{N}[][]{\small{$x_2$}}
	\psfrag{P}[][]{\small{$y_1, y_2$}}
	\psfrag{S}[][]{\small{$\alpha_k$}}
	\psfrag{T}[][]{\small{$\alpha_{k+1}$}}
	\psfrag{X}[][]{\small{$t_k$}}
	\psfrag{Y}[][]{\small{$t_{k+1}$}}
	\psfrag{Z}[][]{\small{$g$}}
	\includegraphics[width=0.85\hsize]{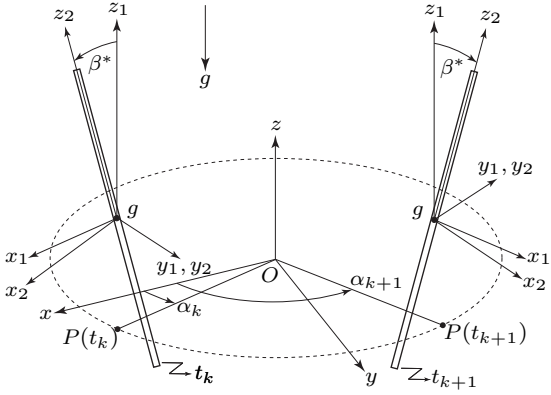}
	\caption{Configurations of the stick at $t_k$ and $t_{k+1}$; the locations of the juggler are denoted by $P(t_k)$ and $P(t_{k+1})$.}
	\label{Fig4}
\end{figure}

We assume that the juggler changes their position intermittently. This position, which is denoted by $P$, is updated immediately prior to application of each impulsive input, \emph{i.e.}, when $\beta = \beta^*$. The point $P$ lies on a circle whose center is the origin of the $xyz$ frame $O$, and $OP$ is parallel to the $x_1$ axis at instants $t = t_k$, $k = 1, 2, \cdots$. The reference frame of the juggler is obtained by rotating the $xyz$ frame intermittently by $\alpha_k$ at $t = t_k^{-}$, $k = 1, 2, \cdots$, about the $z$ axis; this frame remains stationary in the interval $[t_k, t_{k+1}^{-})$. The configurations of the stick and locations of $P$ at instants $t_k$ and $t_{k+1}$ are shown in Fig.\ref{Fig4}.

We denote the position and velocity of the center-of-mass of the stick in the reference frame of the juggler by the vectors $\bar{\vec h}$ and $\bar{\vec v}$. Noting that $\alpha = \alpha_k$ at $t_k^-$ and $\alpha = \alpha_{k+1}$ at $t_{k+1}^-$, \eqref{eq16} can be written as
\begin{subequations} \label{eq28}
\begin{align}
\vec R_z(\alpha_{k+1}) \bar{\vec h}(t_{k+1}^-) &= \vec R_z(\alpha_k) \bar{\vec h}(t_k^-) + \vec R_z(\alpha_k) \bar{\vec v}(t_k^-)\delta_k \nonumber \\ 
&+ \frac{I_k\delta_k}{m} \vec f_k -\frac{1}{2} \begin{bmatrix} 0 \\ 0 \\ g\delta_k^2 \end{bmatrix} \label{eq28a} \\
\vec R_z(\alpha_{k+1}) \bar{\vec v}(t_{k+1}^-) &= \vec R_z(\alpha_k) \bar{\vec v}(t_k^-) + \frac{I_k}{m} \vec f_k - \begin{bmatrix} 0 \\ 0 \\ g\delta_k \end{bmatrix} \label{eq28b}
\end{align}
\end{subequations}

\noindent Premultiplying both sides of \eqref{eq28} by $\vec R_z^T(\alpha_k)$ and using the identity
\begin{equation*}
\vec R_z^T(\alpha_k)\vec R_z(\alpha_{k+1}) = \vec R_z(\Delta\alpha_k), \qquad \Delta\alpha_k \triangleq (\alpha_{k+1} - \alpha_k)
\end{equation*}

\noindent we get
\begin{subequations} \label{eq29}
\begin{align}
\bar{\vec h}(t_{k+1}^-) &= \vec R_z^T(\Delta\alpha_k)\Bigg\lbrace\bar{\vec h}(t_k^-) + \bar{\vec v}(t_k^-)\delta_k \notag \\ 
&+ \frac{I_k\delta_k}{m} \bar{\vec f}_k -\frac{1}{2}  \begin{bmatrix} 0 \\ 0 \\ g\delta_k^2 \end{bmatrix} \Bigg\rbrace \label{eq31a} \\
\bar{\vec v}(t_{k+1}^-) &= \vec R_z^T(\Delta\alpha_k)\Bigg\lbrace \bar{\vec v}(t_k^-) + \frac{I_k}{m} \bar{\vec f}_k - \begin{bmatrix} 0 \\ 0 \\ g\delta_k \end{bmatrix} \Bigg\rbrace \label{eq31b}
\end{align}
\end{subequations}

\noindent where
\begin{equation} \label{eq30}
\begin{split}
\bar{\vec f}_k &\triangleq \vec R_z^T(\alpha_k) \vec f_k = \begin{bmatrix}
- \cos\beta^* \cos\phi_k\\
- \sin\phi_k\\
\sin\beta^*\cos\phi_k
\end{bmatrix} \\
\Delta\alpha_k &= \pi + 2\arctan\!\! \left[\sin\beta^*\! \cos\beta^* 
\frac{\dot\alpha(t_k^-) \!-\! \dfrac{I_k r_k \sin\phi_k}{J \sin\beta^*}}{\dot\beta(t_k^-) \!-\! \dfrac{I_k r_k \cos\phi_k}{J}} \right]
\end{split}
\end{equation}

\noindent The above expression for $\Delta\alpha_k$ was obtained from \eqref{eq19}.\

Based on the description of the reference frame of the juggler, it is clear that the juggler is unaware of the value of $\alpha$ but aware of the values of $\dot\alpha$, $\beta$ and $\dot\beta$. Therefore, in this reference frame, the state vector is
\begin{equation*}
\bar X = \begin{bmatrix} \bar h_x &\bar h_y &\bar h_z &\bar v_x &\bar v_y &\bar v_z &\beta &\dot\alpha &\dot\beta \end{bmatrix}^T
\end{equation*}

\noindent Similar to \eqref{eq26}, the Poincar\'e section in the reference frame of the juggler  is defined as:
\begin{equation} \label{eq31}
\bar S : \{\bar X \in \mathbb{R}^{9} \,\, | \,\, \beta = \beta^* \}
\end{equation}

\noindent A point on $\bar S$ can be described by the vector $\bar Y$, $\bar Y\! \subset \!\bar X$, where
\begin{equation*}
\bar Y = \begin{bmatrix} \bar h_x &\bar h_y &\bar h_z &\bar v_x &\bar v_y &\bar v_z &\dot\alpha &\dot\beta \end{bmatrix}^T
\end{equation*}

\noindent Using \eqref{eq20}, \eqref{eq23}, and \eqref{eq29}, the map $\bar{\mathbb{P}} : \bar S \rightarrow \bar S$ can be expressed in the reference frame of the juggler as
\begin{equation} \label{eq32}
\begin{split}
\bar Y(t_{k+1}^-) &\!= \bar{\mathbb{P}}\left[\bar Y(t_{k}^-), U_k\right] \triangleq \mathcal{R}^T(\Delta\alpha_k)\, \mathcal{F} \!\left[\bar Y(t_k^-), U_k\right] \\
\mathcal{F}\! \left[\bar Y(t_k^-), U_k\right] &\!=\! 
\begingroup
\renewcommand*{\arraystretch}{2.1}
\begin{bmatrix}
\bar h_x(t_k^-) \!+\! \bar v_x(t_k^-)\delta_k \!-\! \dfrac{I_k\delta_k}{m} \cos\beta^* \cos\phi_k \\
\bar h_y(t_k^-) \!+\! \bar v_y(t_k^-)\delta_k \!-\! \dfrac{I_k\delta_k}{m} \sin\phi_k \\
\begin{split}
\bar h_z(t_k^-) \!+\! \bar v_z(t_k^-)\delta_k \!&+\! \dfrac{I_k\delta_k}{m}\sin\beta^*\cos\phi_k\\ &-\dfrac{1}{2}g\delta_k^2
\end{split} \\
\bar v_x(t_k^-) \!-\! \dfrac{I_k}{m} \cos\beta^* \cos\phi_k \\
\bar v_y(t_k^-) \!-\! \dfrac{I_k}{m} \sin\phi_k \\
\bar v_z(t_k^-) \!+\! \dfrac{I_k}{m}\sin\beta^*\cos\phi_k - g\delta_k \\
\dot \alpha(t_k^-) \!-\! \dfrac{I_k r_k \sin\phi_k}{J \sin\beta^*} \\
- \dot \beta(t_k^-) \!+\! \dfrac{I_k r_k \cos\phi_k}{J}
\end{bmatrix}
\endgroup \\
\mathcal{R}(\Delta\alpha_k) &\!\triangleq\! 
\begin{bmatrix} \begin{array}{c|c|c}
\vec R_z(\Delta\alpha_k) & \mathbb{O}_{3 \times 3} & \mathbb{O}_{3 \times 2} \\ \hline
\mathbb{O}_{3 \times 3} & \vec R_z(\Delta\alpha_k) & \mathbb{O}_{3 \times 2} \\ \hline
\mathbb{O}_{2 \times 3} & \mathbb{O}_{2 \times 3} & \mathbb{I}_{2 \times 2}
 \end{array}\end{bmatrix}
\end{split}
\end{equation}

\noindent where $\Delta\alpha_k$ is given by \eqref{eq30}, $\mathbb{O}_{m \times n} \in \mathbb{R}^{m \times n}$ is a matrix of zeros and $\mathbb{I}_{n \times n} \in \mathbb{R}^{n \times n}$ is the identity matrix.\

\begin{remark} \label{rem5}
A sequence of configurations which are rotationally symmetric about the $z$ axis at steady-state (see remark \ref{rem4}) corresponds to a fixed configuration in the reference frame of the juggler. This configuration corresponds to a fixed point of the map $\bar{\mathbb{P}}$ in \eqref{eq32}. Indeed, unlike $\mathbb{P}$, $\bar{\mathbb{P}}$ is not a function of variables in the set $\{h_x, h_y, v_x, v_y, \alpha\}$. The steady-state trajectory of the stick in the reference frame of the juggler is shown in red in Fig.\ref{Fig5}.\
\end{remark}

\section{Control Design for Juggling}\label{sec5}
\subsection{Steady-State Dynamics}\label{sec51}

The objective of juggling the stick between a sequence of rotationally symmetric configurations is equivalent to finding a fixed point of the Poincar\'e Map in \eqref{eq32}, defined in the reference frame of the juggler. A fixed point of the map $\bar{\mathbb{P}}$ is given by the pair $\{\bar Y^*, U^*\}$ which satisfies
\begin{align}
\bar Y^* &= \bar{\mathbb{P}}\left( \bar Y^*, U^* \right)  \label{eq33} \\
\bar Y^* &\triangleq \begin{bmatrix} \bar h_x^* &\bar h_y^* &\bar h_z^*  &\bar v_x^* &\bar v_y^* &\bar v_z^* &\dot\alpha^* & \dot\beta^* \end{bmatrix}^T \notag \\
U^* &\triangleq \begin{bmatrix} I^* &r^* &\phi^* \end{bmatrix}^T\notag
\end{align}

\noindent subject to the constraints that follow from \eqref{eq25} and \eqref{eq30}:
\begin{equation}  \label{eq34}
\delta^* \!=\! \frac{1}{\sqrt{K_1^* + K_2^*}} \left[ \pi - 2\arctan\left( \frac{\sqrt{K_1^* + K_2^*} \cot\beta^*}{\sqrt{K_2^* - K_1^* \cot^2\!\beta^*}} \right) \right]
\end{equation}

\begin{equation} \label{eq35}
\Delta\alpha^* \!=\! \pi + 2\arctan\!\!\left[\sin\beta^*\! \cos\beta^* 
\dfrac{\dot\alpha^* \!-\! \dfrac{I^* r^* \sin\phi^*}{J \sin\beta^*}}{\dot\beta^* \!-\! \dfrac{I^* r^* \cos\phi^*}{J}} \right] 
\end{equation}

\noindent where
\begin{equation*}
\begin{split}
K_1^* &= \sin^4\!\beta^*\left[\dot \alpha^* - \frac{I^* r^* \sin\phi^*}{J \sin\beta^*} \right]^2 \\ 
K_2^* &= \sin^2\!\beta^* \cos^2\!\beta^*\left[\dot \alpha^* - \frac{I^* r^* \sin\phi^*}{J \sin\beta^*} \right]^2 \\
&\qquad\quad + \left[\dot \beta^* - \frac{I^* r^* \cos\phi^*}{J} \right]^2
\end{split}
\end{equation*}

\noindent The relations in \eqref{eq33}, \eqref{eq34} and \eqref{eq35} represent 10 equations in 13 unknowns, namely $\bar h_x^*$, $\bar h_y^*$, $\bar h_z^*$,
$\bar v_x^*$, $\bar v_y^*$, $\bar v_z^*$, $\dot \alpha^*$, $\dot \beta^*$, $I^*$, $r^*$, $\phi^*$, $\delta^*$, and $\Delta\alpha^*$. The unknown $\bar h_z^*$ is eliminated after simplification of \eqref{eq33}; this leaves us with 10 equations in 12 unknowns. We choose the values of $\delta^*$ and $\Delta\alpha^*$ to solve for the remaining ten unknowns:
\begin{equation} \label{eq36}
\begin{aligned}
\bar h_x^* &= \frac{g\delta^{*2} \cot\beta^*}{2(1 - \cos\Delta\alpha^*)},& \bar h_y^* &= 0 \\
\bar v_x^* &= \frac{g\delta^* \cot\beta^*}{2},& \bar v_y^* &= \frac{g\delta^* \cot\beta^* \sin\Delta\alpha^*}{2(1 - \cos\Delta\alpha^*)} \\
\bar v_z^* &= -\frac{1}{2} g\delta^*,& \dot \alpha^* &= \frac{\Psi \sin\Delta\alpha^*}{\delta^* \sin 2\beta^*(1 - \cos\Delta\alpha^*)} \\
\dot \beta^* &= \frac{\Psi}{2 \delta^*},& I^* &= \frac{m g \delta^*}{\sin\beta^*} \\
r^* &= \frac{J \Psi \sin\beta^*}{m g \delta^{*2}},& \phi^* &= 0
\end{aligned}
\end{equation}

\noindent where
\begin{align} \label{eq37}
\Psi &= \frac{2}{\xi} \left[ \pi - 2\arctan\left( \xi \cot\beta^* \right) \right] \\
\xi &\triangleq \sqrt{1 + \sec^2 \beta^* \left[\frac{\sin \Delta\alpha^*}{1 - \cos \Delta\alpha^*}\right]^2} \notag
\end{align}

\noindent The value $\bar h_y^* = 0$ indicates that, at steady-state, the stick lies in the vertical plane containing point $P$ at instants when impulsive inputs are applied; $\phi^* = 0$ indicates that, at steady-state, the impulsive force vector also lies in this plane. The point of application of the impulsive force must lie on the stick, \emph{i.e.}, $0 \leq r^* \leq \ell/2$; this imposes the following constraint on the time of flight:
\begin{equation} \label{eq38}
\delta^* = p\, \delta_{\rm min}, \quad \delta_{\rm min} \triangleq \sqrt{\frac{2 J \Psi \sin\beta^*}{m g \ell}}, \quad p \geq1
\end{equation}

\noindent The above relation requires that $\Psi \geq 0$. Without loss of generality, we assume that $\dot \alpha^* > 0$, \emph{i.e.}, the stick is juggled between a sequence of configurations where one configuration is obtained from the previous by a counterclockwise rotation about the $z$ axis; from the expression of $\dot\alpha^*$ in \eqref{eq36}, the choice of $\Delta\alpha^*$ must satisfy
$\Delta\alpha^* \in (0, \pi]$.

The desired juggling motion of the stick is repetitive in nature and can be described the following hybrid orbit: 
\begin{equation}\label{eq39}
\bar{\mathcal{O}}^* = \{\bar X \in \mathbb{R}^{9} \,\,|\,\, \bar{Y}(t_k^-) = \bar{Y}^*, U_k = U^*\}
\end{equation}  

\noindent In the next subsection, we address the problem of stabilization of the hybrid orbit $\bar{\mathcal{O}}^*$, which is equivalent to stabilization of desired juggling motion.

\subsection{Hybrid Orbit Stabilization}\label{sec52}

In the absence of disturbances, the input vector $U^*$ in (\ref{eq33}) ensures desired juggling motion if  $\bar{Y}(t_k^-) = \bar{Y}^*$. However, if this is not the case, the trajectory of the stick may not converge to the desired orbit $\bar{\mathcal{O}}^*$ with input $U^*$.  It is therefore necessary to determine the stability characteristics of $\bar{\mathcal{O}}^*$. To this end, we define an $\epsilon$-neighborhood of $\bar{\mathcal{O}}^*$ as follows:
\begin{align*}
&N_{\epsilon} = \{\bar X \in \mathbb{R}^{9}  : {\rm{dist}}(\bar X, \bar{\mathcal{O}}^*) < \epsilon\}\cr
&{\rm{dist}}(\bar X, \bar{\mathcal{O}}^*) \triangleq \inf_{Z\in \bar{\mathcal{O}}^*} \|\bar X - Z\|
\end{align*}
\begin{definition}\label{def1}
The orbit $\bar{\mathcal{O}}^*$ in (\ref{eq39}) is
\begin{itemize}
\item stable, if for every $\epsilon >0$, there is a $\delta >0$ such that $\bar X(0) \in N_{\delta} \Rightarrow \bar X(t) \in N_{\epsilon}, \,\, \forall t \geq 0$.
\item asymptotically stable if it is stable and $\delta$ can be chosen such that $\lim_{t \to \infty} {\rm{dist}}(\bar X(t), \bar{\mathcal{O}}^*) = 0$.
\end{itemize}
\end{definition}
\begin{figure}[b!]
\centering
\psfrag{A}[][]{\small{$\bar{S}$}}
\psfrag{C}[][]{\scriptsize{$\bar Y^*$}}
\psfrag{E}[][]{\scriptsize{$\bar Y(t_k^{-})$}}
\psfrag{F}[][]{\scriptsize{$\bar Y(t_k^{+})$}}
\psfrag{G}[][]{\scriptsize{$\bar Y(t_{k+1}^{-})$}}
\psfrag{H}[][]{\scriptsize{$\bar{\mathcal{O}}^*$}}
\includegraphics[width=0.64\linewidth]{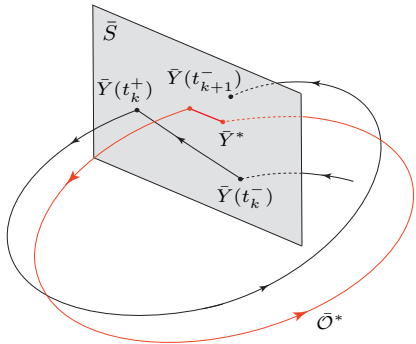}
\caption{Schematic of the ICPM approach to hybrid orbit stabilization. The steady-state orbit of the stick in the reference frame of the juggler is shown in red.}  
\label{Fig5}
\end{figure}
\noindent The hybrid orbit $\bar{\mathcal{O}}^*$ in \eqref{eq39} is asymptotically stable if the fixed point $\bar{Y}^*$ in \eqref{eq33} is asymptotically stable \cite[Th. 1]{kant2022juggling}, which is an abridged version of \cite[Th. 1]{grizzle2001asymptotically} and \cite[Th. 2.1]{nersesov2002generalization}. To stabilize the fixed point $\bar{Y}^*$, we use the ICPM approach \cite{kant2020orbital}, which was developed for continuous orbits, and later extended to hybrid orbits associated with planar stick juggling \cite{kant2022juggling}. The ICPM approach is explained with the help of Fig.\ref{Fig5}. The desired hybrid orbit $\bar{\mathcal{O}}^*$ (shown in red), first intersects the Poincar\'e section $\bar S$ at the fixed point $\bar{Y}^*$. It then undergoes a discontinuous jump due to application of the input $U^*$ before exiting $\bar S$ and evolving continuously. For a trajectory not on $\bar{\mathcal{O}}^*$ (shown in black), the configuration jumps from $\bar{Y}(t_k^-)$ to $\bar{Y}(t_k^+)$ on $\bar S$ due to the application of input $U_k$. Hereafter, the stick undergoes torque free motion under gravity and the next intersection of the continuous-time trajectory with $\bar S$ is denoted by
$\bar{Y}(t^{-}_{k+1})$.

To apply the ICPM approach, we linearize the Poincar\'e map $\bar{\mathbb{P}}$ in (\ref{eq32}) about $\bar{Y} = \bar{Y}^*$ and $U = U^*$ as follows:
\begin{align}\label{eq40}
e(k+1) &= \mathcal{A}e(k) + \mathcal{B}u(k) \\
e(k) &\triangleq \bar{Y}(t^{-}_k) - \bar{Y}^*, \quad u(k) \triangleq U_k - U^* \notag
\end{align}

\noindent where
\begin{equation}\label{eq41}
\begin{split}
\mathcal{A} &\triangleq  \left[\nabla_{\bar{Y}}\mathbb{P}(\bar Y, U)\right]_{\bar Y = \bar Y^*\!\!, \,U = U^*} \\
\mathcal{B} &\triangleq  \left[\nabla_{U}\mathbb{P}(\bar Y, U) \right]_{\bar Y = \bar Y^*\!\!, \,U = U^*}
\end{split}
\end{equation}

\noindent The analytical expressions for $\mathcal{A}$ and $\mathcal{B}$ are not provided here for brevity. From the expressions, it can be verified that the pair $(\mathcal{A}, \mathcal{B})$ is controllable. The hybrid orbit $\bar{\mathcal{O}}^*$ can therefore be stabilized by the following discrete feedback:
\begin{equation}\label{eq42}
u(k) = \mathcal{K}e(k)
\end{equation}

\noindent where the matrix $\mathcal{K}$ is chosen such that the eigenvalues of $(\mathcal{A}+\mathcal{B}\mathcal{K})$ lie inside the unit circle.\

\section{Simulation}\label{sec6}

The physical parameters of the stick in SI units are:
\begin{equation} \label{eq43}
m = 0.1,\quad \ell = 0.5,\quad J = \frac{1}{12}m\ell^2 = 0.0021
\end{equation}
Choosing $\beta^* = \pi/3$ rad, $\delta^* = 0.6$ s, and $\Delta\alpha^* = 2\pi/3$ rad, and using \eqref{eq43} in \eqref{eq36}, we obtain
\begin{equation} \label{eq44}
\begin{aligned}
\bar h_x^* &= 0.6797\ {\rm m},& \bar h_y^* &= 0 \\
\bar v_x^* &= 1.6991\ {\rm m/s},& \bar v_y^* &= 0.9810\ {\rm m/s} \\
\bar v_z^* &= -2.9430\ {\rm m/s},& \dot \alpha^* &= 2.4675\ {\rm rad/s} \\
\dot\beta^* &= 1.8506\ {\rm rad/s},& I^* &= 0.6797\ {\rm Ns} \\
r^* &= 0.0113\ {\rm m},& \phi^* &= 0
\end{aligned}
\end{equation}
The value $\bar h_z^*$ is chosen arbitrarily to be 1.6 m.\

The gain matrix $\mathcal{K}$ in \eqref{eq42}, which is not provided here for brevity, is designed using the LQR method \cite{antsaklis2007linear}; it minimizes the cost functional
\begin{equation*}
\mathcal{J} = \sum_{k = 1}^\infty\left[ e(k)^T Q e(k) + u(k)^T R u(k) \right]
\end{equation*}

\noindent where the $Q$ and $R$ matrices were chosen as
\begin{equation*}
Q = \mathcal{I}_{8 \times 8},\quad R = \mathrm{diag}\begin{bmatrix} 2.0 & 0.5 & 1.0\end{bmatrix}
\end{equation*}

We assume that $\beta(1) = \beta^* = \pi/3$ rad and choose the initial values of the states as
\begin{equation} \label{eq45}
\begin{aligned}
\bar h_x(1) &= 0.9\ {\rm m},& \bar h_y(1) &= -0.2\ {\rm m} \\
\bar h_z(1) &= 1.2\ {\rm m},& \bar v_x(1) &= 1.3\ {\rm m/s} \\
\bar v_y(1) &= 0.2\ {\rm m/s},& \bar v_z(1) &= -1.7\ {\rm m/s} \\
\dot \alpha(1) &= 2.2\ {\rm rad/s},& \dot \beta(1) &= 2.1\ {\rm rad/s}
\end{aligned}
\end{equation}

\begin{figure}[b!]
\centering
\psfrag{A}[][]{\scriptsize{$\bar h_z (k)$ (m)}}
\psfrag{B}[][]{\scriptsize{$\bar h_x (k)$ (m)}}
\psfrag{C}[][]{\scriptsize{$\bar h_y (k)$ (m)}}
\psfrag{D}[][]{\scriptsize{$\bar v_x (k)$ (m/s)}}
\psfrag{E}[][]{\scriptsize{$\bar v_y (k)$ (m/s)}}
\psfrag{F}[][]{\scriptsize{$\bar v_z (k)$ (m/s)}}
\psfrag{G}[][]{\scriptsize{$\dot\alpha (k)$ (rad/s)}}
\psfrag{H}[][]{\scriptsize{$\dot\beta (k)$ (rad/s)}}
\psfrag{I}[][]{\scriptsize{$I_k$ (Ns)}}
\psfrag{J}[][]{\scriptsize{$r_k$ (m)}}
\psfrag{K}[][]{\scriptsize{$\phi_k$ (rad)}}
\psfrag{L}[][]{\scriptsize{$\Delta\alpha_k$ (rad)}}
\psfrag{M}[][]{\scriptsize{$\delta_k$ (s)}}
\psfrag{N}[][]{\small{$k$}}
\psfrag{X}[][]{\small{$\times$}}
\includegraphics[width=1.0\linewidth]{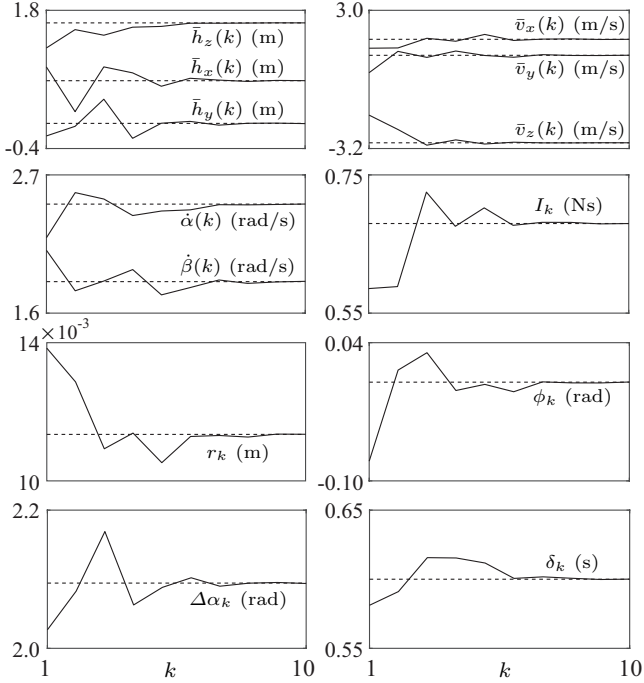}
\caption{Simulation results: the state variables, control inputs, precession $\Delta\alpha_k$, and time of flight $\delta_k$ are shown at time instants $t_k^{-}$, $k = 1, 2, \cdots, 10$, for the initial conditions in \eqref{eq45}.}  
\label{Fig6}
\end{figure}
\begin{figure}[b!]
\centering
\psfrag{A}[][]{\small{$z$}}
\psfrag{B}[][]{\small{$x$}}
\psfrag{C}[][]{\small{$y$}}
\psfrag{D}[][]{\scriptsize{$(t\!=\!0)$}}
\psfrag{E}[][]{\small{$\approx$}}
\includegraphics[width=0.73\linewidth]{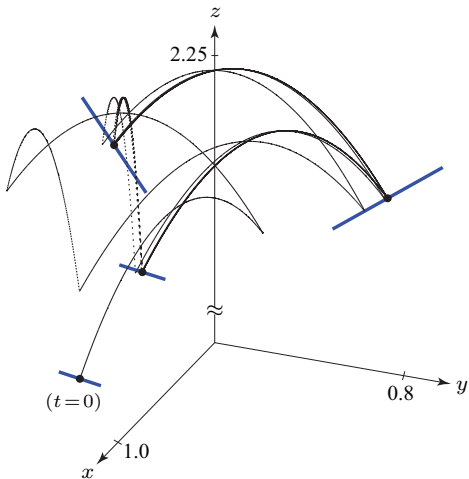}
\caption{Trajectory of the center-of-mass of the stick in the inertial frame; also shown is the stick in its initial configuration ($t = 0$) and three rotationally symmetric steady-state configurations.}  
\label{Fig9}
\end{figure}

\noindent The simulation results are shown in Figs.\ref{Fig6} and \ref{Fig9}. Figure \ref{Fig6} shows that the states and inputs converge to their steady-state values, indicated by dotted lines, in approx. $k = 10$ steps. This validates that desired juggling motion of the stick is achieved through stabilization of the fixed point $\bar Y^*$. The trajectory of the center-of-mass of the stick over approx. $12.02$ s (corresponding to $k = 20$ steps) is shown in Fig.\ref{Fig9}; a video of the juggling motion is uploaded as supplementary material.\

\begin{remark} \label{rem6}
At steady-state, the kinetic energy of the stick is the same before and after the application of each impulsive force; therefore, the impulsive forces do no work. Since the energy of the stick remains constant during the flight phase, the mechanical energy of the stick is conserved at steady-state.
\end{remark}

\begin{figure}[t!]
\centering
\psfrag{A}[][]{\scriptsize{$\bar h_z (k)$ (m)}}
\psfrag{B}[][]{\scriptsize{$\bar h_x (k)$ (m)}}
\psfrag{C}[][]{\scriptsize{$\bar h_y (k)$ (m)}}
\psfrag{D}[][]{\scriptsize{$\bar v_x (k)$ (m/s)}}
\psfrag{E}[][]{\scriptsize{$\bar v_y (k)$ (m/s)}}
\psfrag{F}[][]{\scriptsize{$\bar v_z (k)$ (m/s)}}
\psfrag{G}[][]{\scriptsize{$\dot\alpha (k)$ (rad/s)}}
\psfrag{H}[][]{\scriptsize{$\dot\beta (k)$ (rad/s)}}
\psfrag{I}[][]{\scriptsize{$I_k$ (Ns)}}
\psfrag{J}[][]{\scriptsize{$r_k$ (m)}}
\psfrag{K}[][]{\scriptsize{$\phi_k$ (rad)}}
\psfrag{L}[][]{\scriptsize{$\Delta\alpha_k$ (rad)}}
\psfrag{M}[][]{\scriptsize{$\delta_k$ (s)}}
\psfrag{N}[][]{\small{$k$}}
\psfrag{X}[][]{\small{$\times$}}
\includegraphics[width=1.0\linewidth]{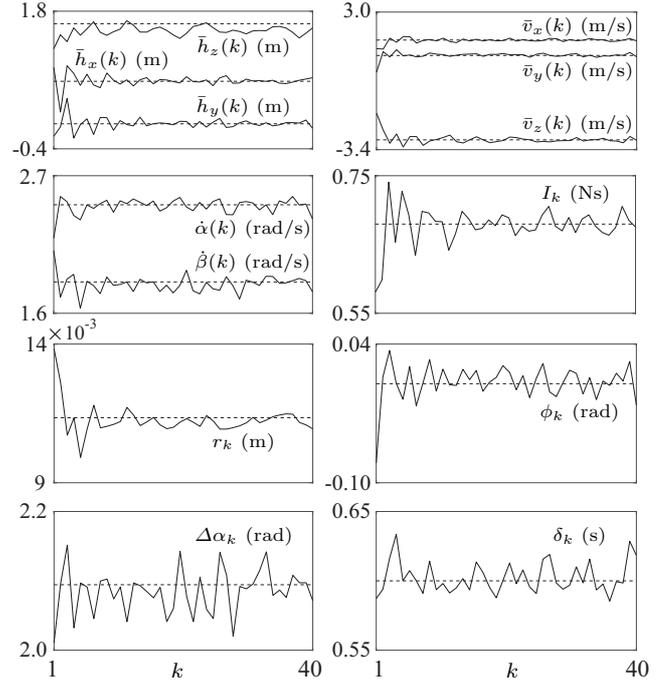}
\caption{Simulation results demonstrating stable behavior in the presence of energy losses: the state variables, control inputs, precession $\Delta\alpha_k$, and time of flight $\delta_k$ are shown at time instants $t_k^{-}$, $k = 1, 2, \cdots, 40$, for the initial conditions in \eqref{eq45}.}  
\label{Fig10}
\end{figure}

To demonstrate robustness of the closed-loop system to losses and uncertainties in the state measurements, we consider a random reduction in the magnitude of the impulsive force $I_k$ in the range 0.0-2.5\% of its computed value, random noise in the position measurements $\bar h_x(k)$, $\bar h_y(k)$, and $\bar h_z(k)$ in the range $\pm 1.0$\%, and random noise in the linear and angular velocity measurements $\bar v_x(k)$, $\bar v_y(k)$, $\bar v_z(k)$, $\dot \alpha(k)$, and $\dot \beta(k)$ in the range $\pm 2.5$\%. For the same parameter values in \eqref{eq43}, steady-state values in \eqref{eq44}, initial conditions in \eqref{eq45}, and gain matrix $\mathcal{K}$, the simulation results are shown in Fig.\ref{Fig10} for $k = 40$ steps. The results of the simulation, which was carried out for a much larger number of steps than shown, indicate that errors in the state variables, control inputs, precession angle, and time of flight remain bounded, implying stable behavior.\

\section{Special Cases}\label{sec7}

\subsection{Planar Juggling}\label{sec71}

The choice $\Delta\alpha^* = \pi$ reduces the problem to that of planar symmetric juggling, which was discussed in \cite{kant2021non, kant2022juggling}. Setting $\Delta\alpha^* = \pi$ in \eqref{eq36}, yields\
\begin{equation} \label{eq46}
\begin{aligned}
\bar h_x^* &= (g\delta^{*2} \cot\beta^*)/4,& \bar h_y^* &= 0 \\
\bar v_x^* &= (g\delta^* \cot\beta^*)/2,& \bar v_y^* &= 0 \\
\bar v_z^* &= - g\delta^*/2,& \dot \alpha^* &= 0 \\
\dot \beta^* &= 2 \beta^*/\delta^*,& I^* &= m g \delta^*/\sin\beta^* \\
r^* &= (4 J \beta^* \sin\beta^*)/m g \delta^{*2} ,& \phi^* &= 0
\end{aligned}
\end{equation}

\noindent which are identical to those derived in \cite{kant2021non, kant2022juggling}. Compared to the results in \cite{kant2021non, kant2022juggling}, the current formulation is more general in that the initial condition can be arbitrary and need not be restricted to lie on the plane of steady juggling.\

\subsection{Steady Precession}\label{sec72}

\begin{figure}[t!]
\centering
\psfrag{A}[][]{\footnotesize{$F^*$}}
\psfrag{B}[][]{\footnotesize{$G$}}
\psfrag{C}[][]{\footnotesize{$\beta^*$}}
\psfrag{D}[][]{\footnotesize{$\bar h_x^*$}}
\psfrag{E}[][]{\footnotesize{$r^*$}}
\psfrag{F}[][]{\footnotesize{$x_1$}}
\psfrag{G}[][]{\footnotesize{$z_2$}}
\psfrag{H}[][]{\footnotesize{$x_2$}}
\psfrag{J}[][]{\footnotesize{$x$}}
\psfrag{K}[][]{\footnotesize{$\alpha$}}
\psfrag{L}[][]{\footnotesize{$z$}}
\psfrag{Z}[][]{\footnotesize{$g$}}
\includegraphics[width=0.55\linewidth]{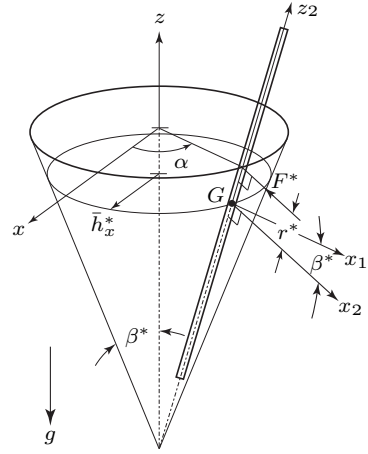}
\caption{Motion of stick in the limiting case of $\Delta\alpha^* \to 0^+$.}  
\label{Fig7}
\end{figure}

In the limiting case where $\Delta\alpha^* \to 0^+$, the surface traced by the motion of the stick defines an inverted cone of semi-vertex angle $\beta^*$ - see Fig.\ref{Fig7}. The discrete set of impulsive forces tend to a continuous force $F^*$ that acts in the direction normal to the cone. The point of application of $F^*$ lies at a distance of $r = r^*$ and traces a circle on the cone. This circle can be viewed as a hoop on which the stick slides without friction and precesses; $F^*$ is the reaction force of the hoop on the stick. The center-of-mass $G$ also traces a circle on the cone; the radius of this circle is $\bar h_x^*$. In the limit, the steady-state time of flight, which satisfies the constraint in \eqref{eq38}, equals
\begin{equation} \label{eq47}
\lim_{\Delta\alpha^* \to 0^+}\delta^* = p\lim_{\Delta\alpha^* \to 0^+}\delta_{\rm min} = 0
\end{equation}

\noindent The above result, which is intuitive, can be obtained from \eqref{eq38} by computing the value of $\Psi$ in \eqref{eq37} in the limit $\Delta\alpha^* \to 0^+$. The continuous force $F^*$ can be expressed by the relation 
\begin{equation} \label{eq48}
F^* = \dfrac{I^*}{\delta^*}
\end{equation}

\noindent since the impulse of the force over the time interval $\delta^*$, which is equal to $F^*\delta^*$, must equal the impulse $I^*$ of the impulsive force at steady state. By replacing the expression for $I^*$ in \eqref{eq36} with $F^*$ in \eqref{eq48} and taking the limit $\Delta\alpha^* \to 0^+$, we get the following values that describe the dynamics of the stick:
\begin{equation} \label{eq49}
\begin{aligned}
\bar h_x^* &= \frac{2p^2 J \sin\beta^* \cos^2\!\beta^*}{m \ell},& \bar h_y^* &= 0 \\
\bar v_x^* &= 0,& \bar v_y^* &= p\sqrt{\frac{2 J g \cos^3\!\beta^*}{m \ell}} \\
\bar v_z^* &= 0,& \dot \alpha^* &= \frac{1}{p}\sqrt{\frac{m g \ell}{2 J \sin^2\!\beta^*\cos\beta^*}} \\
\dot \beta^* &= 0,& F^* &= mg/\sin\beta^* \\
r^* &= \ell/(2p^2), & \phi^* &= 0
\end{aligned}
\end{equation}

\noindent In the above expressions, $p$ is a free parameter. For a specific value of $p$, $p\geq 1$, the stick will exhibit a unique motion. The results in \eqref{eq49} can be independently derived (see Appendix) by considering the continuous-time dynamics of the stick.\

\section{Conclusions}\label{sec8}

Juggling is an example of nonprehensile manipulation using intermittent impulsive forces. An experienced juggler can juggle a stick between desired configurations by intermittently applying impulsive forces while the stick falls freely under gravity. With the motivation of enabling robotic systems to perform complex nonprehensile manipulation tasks that can be performed by humans, the problem of juggling a stick in three-dimensional space is considered in this paper.\

We consider juggling a stick between a sequence of configurations that are rotationally symmetric about the vertical axis. The stick is assumed to be slender and its configuration is described by five generalized coordinates. The location, direction, and magnitude of the impulsive forces applied on the stick are modeled as the control inputs. The control is event-based, with impulsive forces applied when the stick reaches a desired orientation relative to the vertical axis. The hybrid dynamics of the stick is represented using a Poincar\'e map but the sequence of rotationally symmetric configurations do not correspond to a fixed point of this map. To alleviate this problem, the Poincar\'e map is redefined in a rotating reference frame. The trajectory of the stick at steady-state is represented by a hybrid orbit, and stabilizing the desired juggling motion is equivalent to stabilization of the hybrid orbit; this is accomplished through stabilization of the fixed point of the Poincaré map in the rotating reference frame. Using the ICPM approach, the Poincar\'e map is linearized about the fixed point and this results in a controllable linear discrete-time system. Simulation results based on an LQR design are presented to demonstrate stabilization of a desired juggling motion from an arbitrary initial configuration.\

In our approach, the angle of precession of the stick about the vertical axis between consecutive control actions at steady-state can be specified. When this angle equals $\pi$ rad, the model for planar symmetric stick-juggling \cite{kant2021non} is recovered. In the limiting case, when the angle approaches zero, the hybrid dynamics approaches the continuous-time dynamics of steady precession on a hoop. With focus on robotic juggling, our future work will extend the model presented here to include the mechanics of impact, address the motion planning and control problems of the robot end-effector for generating the required impulsive forces, and demonstrate stick-juggling using real hardware.\

\section*{Appendix - Steady Precession of Stick on Hoop}

For continuous motion of the stick, which was described in section \ref{sec72}, we already have $\dot\beta = \dot\beta^* = 0$ and $\phi = \phi^* = 0$. To derive the remaining eight expressions in \eqref{eq49}, we first write the expressions for the force on the stick and its point of application:\
\begin{equation} \label{eq50}
\vec F = F^* \!\begin{bmatrix}
- \cos\alpha \cos\beta^*\\
- \sin\alpha \cos\beta^*\\
\sin\beta^*
\end{bmatrix}, \quad
\vec{r} = r^* \!\begin{bmatrix}
\cos\alpha \sin\beta^* \\
\sin\alpha \sin\beta^* \\
\cos\beta^*
\end{bmatrix}
\end{equation} 

\noindent where $\alpha$ indicates the angle of precession. The motion of the center-of-mass is described by the differential equation\
\begin{equation} \label{eq51}
\dot{\vec h} = \vec v, \qquad \dot{\vec v} = \frac{\vec F}{m} + \begin{bmatrix} 0 & 0 & -g \end{bmatrix}^T
\end{equation}

\noindent with $\vec h$ and $\vec v$ defined in \eqref{eq3}. The angular dynamics is governed by the relation
\begin{equation} \label{eq52}
\dot{\vec H} = \vec r \times \vec F = F^* r^* \begin{bmatrix}
\quad \sin\alpha \\
-\cos\alpha \\
0
\end{bmatrix}
\end{equation}

\noindent where $\vec F$ and $\vec r$ are given in \eqref{eq50}. The last equation implies $H_z$ is constant. Using \eqref{eq6} and \eqref{eq7}, we get
\begin{equation} \label{eq53}
H_z = J \dot\alpha \sin^2\beta^* = {\rm constant} \quad \Rightarrow \quad \dot\alpha = {\rm constant}
\end{equation}

\noindent By denoting the constant value of $\dot\alpha$ by $\dot\alpha^*$ and comparing $\dot H_x$ and $\dot H_y$ in \eqref{eq52} with the derivatives of $H_x$ and $H_y$ in \eqref{eq6}, we get
\begin{equation} \label{eq54}
J \dot\alpha^{*2} \sin\beta^* \cos\beta^* = F^* r^*
\end{equation}

\noindent The motion of the center-of-mass in \eqref{eq51} can be expressed in the reference frame of the juggler by multiplying both sides of both equations by
$\vec R_z^T(\alpha)$; this yields
\begin{equation} \label{eq55}
\dot{\bar{\vec h}} = \bar{\vec v}, \qquad \dot{\bar{\vec v}} = \frac{F^*}{m} \begin{bmatrix} -\cos\beta^* \\0 \\\sin\beta^* \end{bmatrix} - \begin{bmatrix} 0 \\0 \\g \end{bmatrix}
\end{equation}

\noindent Since $\bar h_z$ is constant, we have
\begin{equation} \label{eq56}
\bar v_z = \bar v_z^* = 0 \quad \Rightarrow \quad \dot{\bar v}_z = 0
\end{equation}

\noindent Using $\dot{\bar v}_z = 0$, we get from \eqref{eq55}
\begin{equation} \label{eq57}
F^* = mg/\sin\beta^*
\end{equation}

\noindent We now make the choice
\begin{equation} \label{eq58}
r^* = \ell/(2p^2), \qquad p \geq 1
\end{equation}

\noindent such that the point of application of the force satisfies $r = r^*$, $0 \leq r^* \leq \ell/2$. Substituting \eqref{eq57} and \eqref{eq58} in \eqref{eq54}, we obtain
\begin{equation} \label{eq59}
\dot \alpha^* = \frac{1}{p}\sqrt{\frac{m g \ell}{2 J \sin^2\beta^* \cos\beta^*}}
\end{equation}

\noindent The center-of-mass of the stick undergoes uniform circular motion with a radius $\bar h_x = \bar h_x^*$; therefore, we have 
\begin{equation} \label{eq60}
\bar v_x^* = 0, \qquad \dot{\bar v}_x^* = -\dot\alpha^{*2}\, \bar h_x^*
\end{equation}

\noindent Comparing the expressions for $\dot{\bar v}_x^*$ in \eqref{eq55} and \eqref{eq60} and substituting the value of $\dot\alpha^*$ in \eqref{eq59}, we get
\begin{equation} \label{eq61}
\bar h_x^* = \frac{2p^2 J \sin\beta^* \cos^2\beta^*}{m \ell}
\end{equation}

\noindent For the motion of the center-of-mass of the stick, we can also write
\begin{equation} \label{eq62}
\bar h_y^* = 0, \qquad \bar v_y^* = \dot\alpha^* \bar h_x^*
\end{equation}

\noindent Substituting \eqref{eq59} and \eqref{eq61} in the expressions for $v_y^*$, we get
\begin{equation} \label{eq63}
\bar v_y^* = p\sqrt{\frac{2 J g \cos^3\beta^*}{m \ell}}
\end{equation}

\noindent Equations \eqref{eq56}-\eqref{eq63} verify the expressions in \eqref{eq49}.\

\vspace{-0.10in}
\section*{Acknowledgements}
The authors acknowledge the support provided by the National Science Foundation, Grant CMMI-2043464.

\vspace{-0.10in}
\section*{Conflict of Interest}
The authors declare that they do not have any conflict of interest.\

\vspace{-0.10in}
\section*{Data Availability Statement}
The datasets generated during and/or analyzed during the current study are available from the corresponding author on reasonable request.\

\balance
\bibliographystyle{plain}
\bibliography{ref}

\end{document}